# Prediction of the Economic Behavior of Fishery Biotechnology Companies Based on machine learning-based deep metacellular automata

Liguo Chen, University of Illinois at Urbana-Champaign, Urbana, IL 61801, USA；
Hongyang Hua, Beijing No. 2 High School International Department, Beijing, 100010；
Xinyue Luo, Guangdong University of Finance & Economics, Guangzhou, 510320, China；
Guoli Xu, New York University, New York, NY 10012, USA；
Xu Yan*, The University of Nottingham, Nottingham, NG7 2RD, UK
*Corresponding author, Email: shyxy2@nottingham.ac.uk

**Abstract:** Ocean warming significantly affects the fishing industry, with species like Scottish herring and mackerel migrating northwards. Our research, a fusion of artificial intelligence, data science, and operations research, addresses this crisis. Using Long Short Term Memory networks, we forecast sea surface temperatures (SST) and model fish migratory patterns with Enhanced Cellular Automata. A corrective factor within our model adjusts for human impact on SST, guiding diverse mitigation scenarios. We apply operational research to strategize responses, including the modernization of fishing vessels as a less costly alternative to relocation. Our data-driven approach, suggesting fleet modernization, strategic relocation, and product diversification, offers an effective approach to mitigating the threats to the ocean warming phenomenon.

## 1. Introduction

The phenomenon of global ocean warming presents a critical challenge to marine ecosystems, significantly impacting fisheries that form the economic backbone of coastal communities worldwide. Notably, the migration of species such as Scottish herring and mackerel towards cooler northern waters exemplifies the urgent need for adaptive strategies to sustain these vital industries [1]. This shifting marine biodiversity, driven by rising sea temperatures, necessitates a multifaceted research approach that combines environmental science with cutting-edge technology to develop predictive models and practical solutions. Recent advancements in artificial intelligence (AI) and machine learning, particularly in Long Short Term Memory (LSTM) networks and Enhanced Cellular Automata (ECA) models, offer promising tools for forecasting sea surface temperatures (SST) and simulating the complex migration patterns of fish in response to environmental changes [4,5]. These technologies draw from a wide range of applications, from healthcare, where graph convolutional networks (GCN) have revolutionized diagnostics [2,3], to dynamic object tracking in computer vision [4], demonstrating the versatility and power of AI in tackling multifaceted problems. The application of AI extends beyond predictive modeling to include monitoring and diagnostics, with significant implications for environmental conservation and management. For instance, the use of GCNs for Alzheimer's disease diagnosis [6] mirrors potential applications in tracking marine species' migrations [7,8]. Furthermore, dynamic convolutions have been applied to fast object tracking [9], suggesting analogous methods could enhance real-time monitoring of fish populations [10,11]. Enhanced Cellular Automata (ECA) models, inspired by advancements in machine learning, provide a robust framework for simulating environmental and biological processes. This approach aligns with the development of multi-source review-based models [12,13] and multi-scale graph convolutional networks [14,15], showcasing the integration of diverse data sources and scales to improve model accuracy and applicability. The inclusion of technologies such as enhanced vision transformers [16] and 3D point cloud classification [17] in environmental research highlights the cross-disciplinary potential of AI to address complex ecological challenges. Similarly, the optimization of decision-making processes through integrative paradigms combining xgboost and xdeepfm algorithms [18,19] illustrates the transformative impact of AI on strategic planning and policy development in the face of climate change. Our study draws upon this rich body of work to propose a comprehensive model that addresses the impacts of global ocean warming on fisheries. By leveraging insights from software-hardware co-design for efficient model inference [20] and employing self-supervised learning techniques for 3D scene flow estimation [21, 22, 23, 24], we aim to develop a holistic framework that not only predicts environmental changes but also provides actionable strategies for adaptation and mitigation.

In conclusion, the integration of AI and machine learning with environmental science offers a promising avenue for addressing the challenges posed by global ocean warming. This interdisciplinary approach, informed by a wide array of research findings [25, 26, 27], provides the foundation for our study, which seeks to mitigate the adverse effects of climate change on marine ecosystems and the fisheries industry. There are also several image-based recognition work [28, 29, 30]. Through the development of advanced predictive models and the application of innovative technologies, we aspire to contribute to the sustainability and resilience of fisheries in a rapidly changing world.

## 2. Ocean Temperature Prediction Model

The target that simulates the fish migration route due to ocean warming drives us to predict temperature over the next 50 years in the first place. Since people have paid considerable attention to global warming, we find abounding instrumental literature adopted time domain feature series and Long Short Term Memory (LSTM) network to anticipate ocean temperature. In this paper, we regenerate this model to tackle our first problem.

### 2.1 Time-domain Feature Series

A Time Series is a sequence of data recorded in time order and reveals the development of things over a period. On account of its definition, Sea Surface Temperature (SST) is a time sequence. Since we intend to forecast its change in the next 50 years, we extract its Time Domain Feature (TDF) instead of morphological or model features to elicit global structural feature of SST.

When we make the foresight for a specific time $t$, partial data reported exactly right before $t$ may be enough. Hence, for an SST time sequence $T = \{t_1, t_2, \ldots, t_n\}$, we define that the previous sequence we require has the length $l$ ($n$ is far greater than $l$), which marks the number of earlier samples. Apart from this, we introduce variables $l_x$ and $l_y$ as well. The former indicates the number of consecutive months adjacent to the predicted month and refers to the temperature trend and the auto-correlation. As for the latter, it represents the number of successive months alongside that month in a diverse year and implies the seasonal temperature and its tendency. Based on these parameters, we take reckon SST with its TDF and deduce the algorithm:

### 2.2 LSTM Neural Network

LSTM is a type of gated Recurrent Neural Network (RNN). Compared with RNN, the gated structure of LSTM can eliminate gradient disappearance or explosion by modifying self-circulation weight. The most prevalent LSTM has the structure shown in Figure 1. We consider the hidden layer as a memory cell, which has various states as time passes. Here the net depth suggests the time spans.

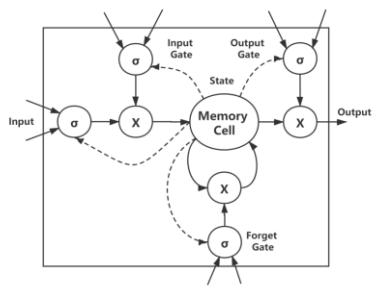

Figure 1: the structure of LSTM memory cell

As illustrated, the input gate is required to determine whether the input can be accumulated to the present state. The memory unit can achieve linear self-circulation while the forget gate alters its weight. Furthermore, outputs are subject to the output gate, which can inhibit it. All gated structure implements nonlinear transformation based on sigmoid function i.e,

$$\sigma(x) = \frac{1}{1+e^{-x}} \tag{3-1}$$

Therefore, the calculation for LSTM network consists of the following equations:

$$i_t = \sigma(W^i H + b^i) \tag{3-2}$$

$$f_t = \sigma(W^f H + b^f) \tag{3-3}$$

$$o_t = \sigma(W^o H + b^o) \tag{3-4}$$

$$c_t = \tanh(W^c H + b^c) \tag{3-5}$$

$$m_t = f_t \ e \ m_{t-1} + i_t \ e \ c_t \tag{3-6}$$

$$h_t = \tanh(o_t \ e \ m_t) \tag{3-7}$$

Where $i_t$, $f_t$ and $o_t$ indicate values of input gate, forget gate and output gate respectively. $c_t$ denotes the new state of a memory cell, while $m_t$ means its ultimate state. Weight matrices are described as $W^i$, $W^f$, $W^o$ and $W^c$ and their corresponding biases are $b^i$, $b^o$, $b^o$ and $b^c$. Moreover, $H$ connects the new input $x_t$ and the previous hidden vector $h_{t-1}$, and equals to $[Ix_t \ h_{t-1}]^T$. Eventually, we get the output $h_t$.

## 2.3 Prediction Model

Combining two structures together, we construct the model as in Figure 2. Back Propagation Through Time (BPTT) algorithm trains the network. We set the mean square error as the loss function.

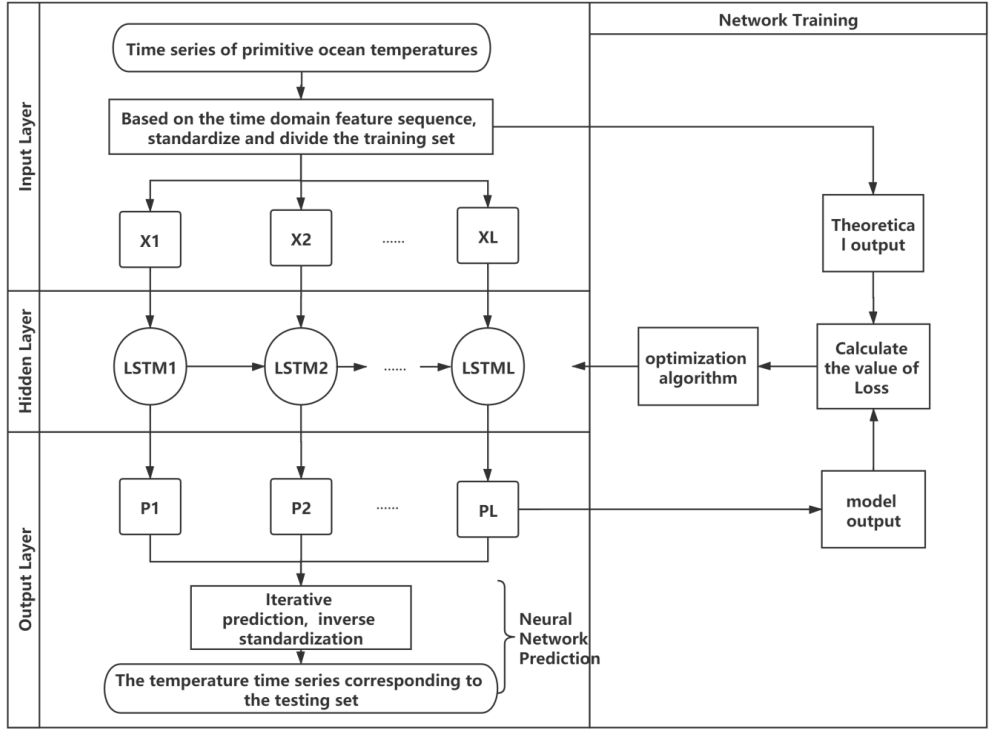

Figure 2: the framework of our prediction model based on LSTM

## 3. Enhanced Cellular Automata

Cellular Automata (CA) performs well in simulating animals movement. However, classical models cannot address our problems, because the number of fish may be not identical in distinct areas, which should be displayed in our model. Therefore, we improve CA by increasing the number of cells in one grid, endowing grid with states and making local rule based on actual fish features. We name the novel model Enhanced Cellular Automata (ECA).

## 3.1 Fishes Living Characteristics

We learn fishes features from Ocean Biogeographic Information System (OBIS), which is an global open-access data associated with marine biodiversity. There are 855,545 records for clupea harengus and 119,487 records for scomber scombrus. We extract temperature data to derive their living conditions.

Noticeably, all samples look normally distributed. In order to checkout our presumption, we utilize Quantile Quantile Plot (QQP) by SPSS. Knowing the distribution, we attain their mean $\mu$ and standard deviation $\sigma$. Therefore, we hold the view that the most suitable SST $T_b$ for fish equals to mean $\mu$ and livable sea temperature scope $T_s$ is $[\mu-2\sigma, \mu+2\sigma]$, which contains 95.45% samples. We list their living characteristics in Table 1 and are identified by subscripts.

Table 1: fish living characteristics

| The sort of fish | Mean $\mu$ (°C) | standard deviation $\sigma$ (°C) | The most suitable SST $T_b$ (°C) | Livable sea temperature scope $T_s$ (°C) |
|---|---|---|---|---|
| Clupea Harengus | $\mu_{ch} \approx 10.397$ | $\sigma_{ch} \approx 0.010$ | $T_{bch} = 10.397$ | $T_{sch} = [10.377, 10.417]$ |
| Scomber Scombrus | $\mu_{ss} \approx 10.304$ | $\sigma_{ss} \approx 0.119$ | $T_{bss} = 10.304$ | $T_{sss} = [10.066, 10.542]$ |

## 3.2 Model Building

Whereas we make some changes, primary definitions are retained:
1) **Cell and cellular space:** Cells point out fish positions and distribute among the sea around Scotland, which is a two-dimensional cellular space after dividing into square grids. In a grid, there are at most 3 cells, each of which comprises 10000 fish.
2) **State:** In the traditional model, the state is merely for cells, nonetheless, we vest grid with it as well in this essay. The grid state $G$ includes four attribute-- $i^{th}$ row, $j^{th}$ column, SST $T_{ij}$ and the number of cells $S_{ij} (S_{ij} \leq 3)$. Besides, we define $n$ to signify the year. For instance, in 2020 $n$ equals to 1 and $n$ will be equivalent to 50 in 2069. Accordingly, SST in the $n^{th}$ year, SST in the $i^{th}$ row and $j^{th}$ column of the grid is $T_{ij}^n$, which is the average temperature of four vertexes. $T^n$ is a SST matrix of the $n^{th}$ year. Once the sea temperature changes, fish moves. For this reason, the cell state can be described as "steady" or "unsteady" according to current living temperature $T_{ij-k}^n$ (where $k$ is the $k^{th}$ cell in this grid and is less than $S_{ij}$). If "steady", fish do not leave; if not, fish relocate on account of local rule. An example is presented with Figure 3 to make a better understanding. In a certain year $n$, a group of fish (cell) stays in a grid whose SST is proper and is in "steady" state. In the next year, SST climbs, as a result, fish feel less cozy and turns to "unsteady" state. This state forces fish to swim to a nearby grid whose SST is the closest to the most suitable SST $T_b$ (movement trend). Ultimately, this group of fish regains "steady" state with the new location.

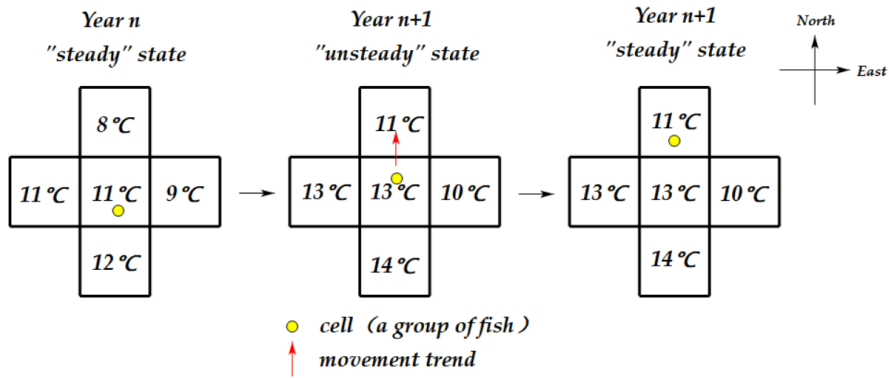

Figure 3: an example to help understand cell states

3) **Neighbor:** In view of fish swimming pace is generally limited, we adopt Von Neumann type to

define cell neighbor.
4) **Local rule:** The normal distribution acknowledged in section 3.1 is the key to determine the probability towards different diverse directions. The probability density function for normal distribution is

$$f(T) = \frac{1}{\sqrt{2\pi}\sigma} e^{-\frac{(T-\mu)^2}{2\sigma^2}} \qquad (4\text{-}1)$$

And the probability distribution function is

$$F(T) = \int_{-\infty}^{T} f(u)\,du \qquad (4\text{-}2)$$

Thus, the transition probability towards one of neighbors is

$$p_t = \begin{cases} \dfrac{\int_{T_{ij}^n}^{T_b} f(u)\,du}{\dfrac{1}{2}\int_{-\infty}^{T} f(u)\,du}, & \text{for } T_{ij}^n \leq T_b \\[2ex] \dfrac{\int_{T_b}^{T_{ij}^n} f(u)\,du}{\dfrac{1}{2}\int_{-\infty}^{T} f(u)\,du}, & \text{for } T_{ij}^n > T_b \end{cases} \qquad (4\text{-}3)$$

For example, if $T_{ij}^n$ is less than $T_b$ in a grid, then the fish may move to adjacent coziest grid with probability $p_t$ till $T_{ij}^n \in T_s$. Once $T_{ij}^n$ is greater than $T_b$ in a grid, the fish may swim towards its neighbor whose SST is closest to $T_b$ with probability $P_t$ till $T_{ij}^n \in T_s$. The shaded part in Figure 6 is the numerator of $p_t$.

## 4. Predicted Locations and Elapsed Time

The crux of the first two tasks is SST in the future. Accordingly, we make use of the Ocean Temperature Prediction Model at the beginning. Once we acquire fish route maps from ECA, we estimate its longitude and latitude according to the number of grids. Predicted SST and locations will be shown in identical grids to improve the clarity of the whole article. As for the second problem, we introduce a correction coefficient to alter the change speed of SST. Based on our understanding of this problem mentioned in the previous section, we deliver a method to evaluate the elapsed time.

### 4.1 Predicted SST

We download SST data over 1870-2019 from Met Office, the national meteorological agency of the United Kingdom, and only use August data because it is the fishing time. MATLAB assists us to handle the data and results are drawn vividly in Figure 5.

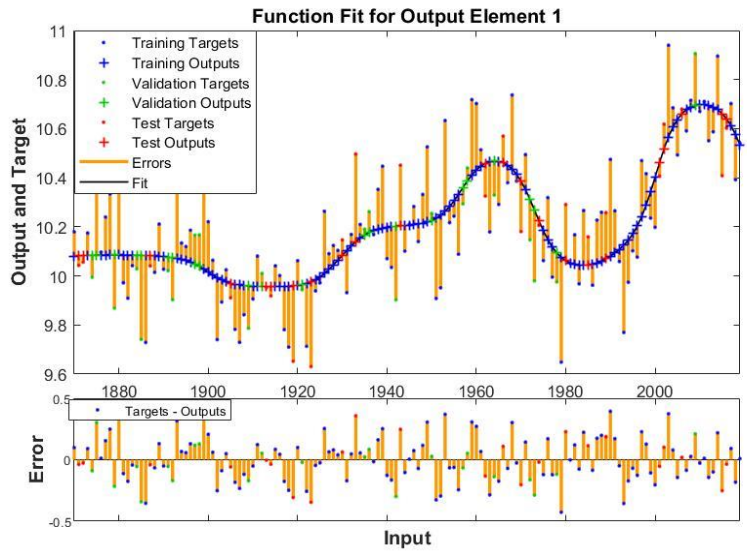

Figure 4: fitted curve

We choose a point at latitude 60.5 degrees north and longitude 7.5 degrees east as an example and present its fitted and predicted consequences in Figure 4. Apparently, SST fluctuated wildly, so that our curve cannot always match very well and predicted curve is almost a "straight line". Fortunately, under our control, the greatest error is merely $0.5^{\circ}C$, which met our expectations.

Based on assumption 8 and 9 in section 2.2, we grid the digital map and set that each grid has side length $1^{\circ}$ (about 111.1km). Each grid is a LSTM block and takes the average SST of four vertexes as its SST. As a cornerstone for the first problem, we estimate SST for cell space i.e, the gridding digital map every 10 years over the next 50 years in Figure 5.

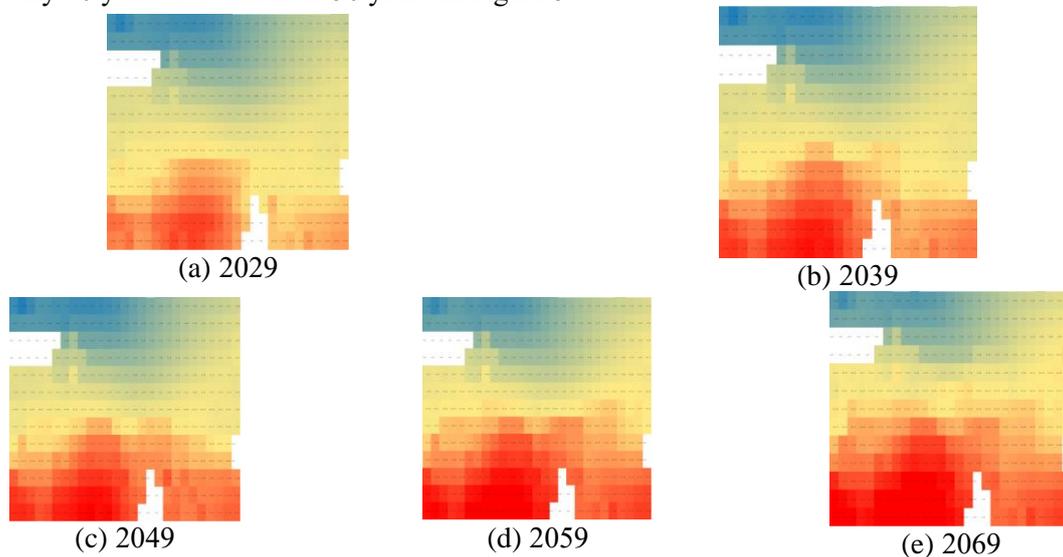

(a) 2029 (b) 2039
(c) 2049 (d) 2059 (e) 2069

Figure 5: predicted SST map

The coldest waters ($2^{\circ}C$) are painted blue while the warmest ones ($12^{\circ}C$) are red. Additionally, white blocks are for lands and the yellowish have the greatest suitable SST. The scopes of longitude and latitude are $[-20.5, 55.0]$ and $[55.5, 67.5]$ (The precision of coordinates is 0.5 degree). Scotland locates at the bottom right. Iceland settles on the left and Norway situates on the right. Every map consists of $27 \times 13$ blocks.

## 4.2 Elapsed Time

SST ascended fast when the Industrial Revolution broke out in 1960s. Then, it dropped briskly in the 1970s, when people developed environmental awareness. This trend suggests that human activities have a great impact on SST. In the predicted part, the "straight line" seems to illustrate the average trend of SST. To simplify this problem, we consider all SST of waters near Scotland, i.e, the North Sea, rise at rate $k$. We then introduce the correction factor $\alpha$, which takes humankind's activities into account, to modify SST changing pace.

## 5. Impacts on Economic Behavior of Fishing Industry

Consider, firstly, the economic implications imparted by the augmentation of fishing vessels, which serves as an economically feasible practice when fish migration patterns remain in proximity. This denotes a form of operational efficiency - forcing cost savings through limiting fuel consumption and reducing seafaring time. By extension, minimal time spent at sea curtails crew-related expenditures for the smaller fishing outfits, contributing to a positive domino effect on their financial statements. Secondly, an intriguing trick of biological vagary - the latency in marine life movement - paves the way for an economic shift. This compels the industry to condone the potential replacement of smaller companies by those with robust financial shoulders capable of enduring unproductive spells. Firms with liquidity resilience and market leverage can capitalize on these pauses in marine activity as an opportunity to amass a greater market share. This, in turn, may mitigate the risk of entire industry decimation during fallow periods, preserving the community's economic fabric reliant on this sector. Finally, the scenario where the lion's share of fish migrates to Norwegian waters calls for strategic redirection. The shifts in migration impose a non-negotiable economic burden on small fishing enterprises, compelling them to reconfigure their operational strategy. Rather than simply cease operations, they are prescribed to scout alternative income-generating tactics. This could range from focusing on niche local species, venturing into aquaculture, or even pivoting towards tourism. Through these economic adaptations, small fishing companies can keep afloat, demonstrating resilience, flexibility and enduring commitment to their economic survival in the volatile marine economy. Thus, rather than being mere victims of natural phenomena, fishing firms become agents in their own economic fate - using their agility, adaptability, and entrepreneurial spirit not just to survive but also to thrive in a world of continuous change.

## 6. Conclusion

We employ LSTM network based on time series which shows time-domain features to predict SST of Scottish waters over the next 50 years. Regarding the living conditions of fish, we introduce Enhanced Cellular Automata to trace fish movement paths. In this essay, we deliver the predicted consequences of 5 years with proper graph. We initially applied a static model to determine the number of fish in a block of water. Founded on the result, another static model derives the expression of profit. Depending on the last dynamic model for estimating profit, we conclude that small fishing companies should make some changes. Buying refrigerators is one of the most economical ways to upgrade fishing boats. As for the territorial issues, we analyze the sensitivity of the profit to the sailing distance and the sailing time in detail. The integrated conclusion is: 1) Improving vessels works when fish swim not too far. 2) when fish moves after a period, the most economical way is to replace firms. 3)When most fish enter Norway, it is better for small fishing companies to turn to another profit method. Moreover, other machine-based tools can be also leveraged to facilitate this process [31-44].